\documentclass[pra,reprint,showpacs,raggedbottom,nofootinbib]{revtex4-1}
\usepackage[active]{srcltx} 
\usepackage{bm,dcolumn}
\usepackage{amsmath,amsfonts,amssymb,amsthm,amscd} 
\usepackage{graphicx}
\usepackage{epstopdf}

\begin{document}
\sloppy
\title{Magic wavelengths for the  $5s-18s$ transition in rubidium}
\author{E. A. Goldschmidt}
\affiliation{Joint Quantum Institute, National Institute of Standards and Technology and the University of Maryland, Gaithersburg, Maryland 20899 USA}
\author{D. G. Norris}
\affiliation{Joint Quantum Institute, National Institute of Standards and Technology and the University of Maryland, Gaithersburg, Maryland 20899 USA}
\author{S. B. Koller}
\altaffiliation[Present address: ]{Optical Lattice Clocks Working 
Group, Physikalisch Technische Bundesanstalt, 38116 Braunschweig, Germany}
\affiliation{Joint Quantum Institute, National Institute of Standards and Technology and the University of Maryland, Gaithersburg, Maryland 20899 USA}
\author{R. Wyllie}
\altaffiliation[Present address: ]{Quantum Systems Division, Georgia Tech Research Institute, Atlanta, GA 30318}
\affiliation{Joint Quantum Institute, National Institute of Standards and Technology and the University of Maryland, Gaithersburg, Maryland 20899 USA}
\author{R. C. Brown}
\affiliation{Joint Quantum Institute, National Institute of Standards and Technology and the University of Maryland, Gaithersburg, Maryland 20899 USA}
\author{J. V. Porto}
\email{porto@umd.edu}
\affiliation{Joint Quantum Institute, National Institute of Standards and Technology and the University of Maryland, Gaithersburg, Maryland 20899 USA}
\author{U. I. Safronova}
\affiliation{Physics Department, University of Nevada, Reno, Nevada 89557 USA}
\author{M. S. Safronova}
\affiliation{Department of Physics and Astronomy, University of Delaware, Newark, Delaware 19716 USA}
\affiliation{Joint Quantum Institute, National Institute of Standards and Technology and the University of Maryland, Gaithersburg, Maryland 20899 USA}

\pacs{32.80.Rm, 37.10.Gh, 31.15.ap}

\begin{abstract}
Magic wavelengths, for which there is no differential ac Stark shift for the ground and excited state of the atom, allow trapping of excited Rydberg atoms without broadening the optical transition. This is an important tool for implementing quantum gates and other quantum information protocols with Rydberg atoms, and reliable theoretical methods to find such magic wavelengths are thus extremely useful. We use a high-precision all-order method to calculate magic wavelengths for the $5s-18s$ transition of rubidium, and compare the calculation to experiment by measuring the light shift for atoms held in an optical dipole trap at a range of wavelengths near a calculated magic value.
\end{abstract}

\maketitle
\section{Introduction}
The concept of a magic wavelength, $\lambda_{\rm{magic}}$, at which two atomic states experience the same ac Stark shift in a light field, was first proposed in Refs.~\cite{KatIdoKuw99,YeVerKim99} for applications in optical atomic clocks. The subject of magic wavelengths has since become of great interest owing to a variety of other applications including laser cooling of fermionic ultracold gases with high phase-space densities \cite{DuaHarHit11}, trapping and controlling atoms in high-Q cavities in the strong coupling regime \cite{MckBucBoo03}, and the implementation of Rydberg-based quantum computing protocols with neutral atoms~\cite{SafWilCla03,Saffman,Piotrowicz2013}. Extensions of the magic wavelength idea include the use of bichromatic trapping to cancel the ac Stark shift~\cite{AroSafCla10} and the use of ``tune-out'' wavelengths where the ac Stark shift is zero for a particular state~\cite{LebThy07,AroSafCla11,HerVaiLi12,Cronin}. Recently, a magic wavelength optical lattice for a Rydberg transition in rubidium was used in a demonstration of light-atom entanglement~\cite{Kuzmich}. 

Such a variety of applications requires the development of theoretical and experimental methods to reliably evaluate various magic wavelengths. While most previous applications involved relatively low-lying states, development of fault-tolerant Rydberg gate schemes calls for accurate prediction and measurement of magic wavelengths for highly-excited states, which is the subject of the present work. Specifically, we theoretically determine $5s-18s$ magic wavelengths near the $18s-6p$ resonances in $^{87}$Rb for Rydberg-based quantum information applications and experimentally measure the magic wavelength near the $18s-6p_{3/2}$ resonance in a crossed-beam optical dipole trap operating near 1064~nm, in the range of standard commercial high-power fiber amplifiers used for optical trapping. We find that the experimentally measured value of 1063.529(4)~nm differs from the theoretically determined value of 1063.514(4)~nm by $\approx2.8\sigma$. Extensive tests of the accuracy of the theoretical calculations and studies of the statistical and systematic uncertainties of the experiment, described later in text, were carried out. We note that the theoretical value depends on the experimentally determined energy levels as described in Section II. Therefore, we also experimentally determine the absolute frequency of the $6p_{3/2}-18s$ transition and obtain a value, 1063.6278(2)~nm, in good agreement with the value extracted from previous measurements, 1063.627(1)~nm~\cite{Sansonetti1985,NIST_ASD}. 

Theoretical determination of magic wavelengths involves calculation of the frequency-dependent polarizabilities of the two states to find their crossing points. A high-precision all-order method was very successful in accurate calculation of atomic polarizabilities for low-lying states~\cite{SafJoh08,AroSafCla07,Li}. However, this approach requires construction of a finite basis set in a spherical cavity of sufficient size to accommodate the relevant electronic orbitals. Thus, owing to the required drastic increase in the cavity size, this approach was previously considered not to be applicable for highly-excited states. In this work, we demonstrate, for the first time, that the resulting problems can be overcome, which significantly expands the applicability of the all-order method. Extensive tests of the numerical stability of the calculations are carried out.\footnote{Unless stated otherwise, for the theoretical calculation we use the conventional system of atomic units, a.u., in which $e, m_{\rm e}$, $4\pi \epsilon_0$ and the reduced Planck constant $\hbar$ have the numerical value 1. The atomic units for $\alpha$ can be converted to SI units via $\alpha/h$~[Hz/(V/m)$^2$]=2.48832$\times10^{-8}\alpha$~[a.u.], where the conversion coefficient is $4\pi \epsilon_0 a^3_0/h$ and the Planck constant $h$ is factored out.}

\section{Theoretical methods}
The frequency-dependent scalar polarizability, $\alpha(\omega)$, of Rb $ns$ states may be separated into a core polarizability and a (dominant) contribution from the valence electron, $\alpha^v(\omega)$. The core polarizability depends weakly on $\omega$ for the frequencies treated here and is approximated by its dc value calculated in the random-phase approximation (RPA)~\cite{MitSafCla10}. The valence part of $\alpha(\omega)$ is evaluated as the sum over intermediate $k$ states allowed by the electric-dipole transition rules~\cite{MitSafCla10}
\begin{equation}
    \alpha^{v}(\omega)=\frac{2}{3(2j_v+1)}\sum_k\frac{{\left\langle k\left\|d\right\|v\right\rangle}^2(E_k-E_v)}{(E_k-E_v)^2-\omega^2}, \label{eq-1}
\end{equation}
where $j_v=1/2$ for $ns$ states, ${\left\langle k\left\|d\right\|v\right\rangle}$ are the reduced electric-dipole matrix elements, and $\omega$ is assumed to be at least several linewidths detuned from the corresponding transition. The calculation of polarizability reduces to the calculation of energies, electric-dipole matrix elements, and evaluation of a small Eq.~(\ref{eq-1}) sum remainder for very large $k$.

In the present work, we are interested in the 18s ac polarizability near 1064~nm, which corresponds to the $18s-6p_{3/2}$ resonance. Therefore, the $18s-6p$, $18s-19p$ $18s-18p$, $18s-17p$, and $18s-16p$ transitions are expected to give large and partially canceling contributions to the polarizability. Accurate representation of such highly-excited states with high-precision all-order methodology is the biggest challenge of the present calculation. First, we carried out extensive numerical tests to ensure that a 500 point grid is sufficient for accurate integration of the corresponding matrix elements at the Dirac-Fock (DF) level. We have compared the values of the $18s-6p$ and $18s-18p$ DF dipole matrix elements integrated on the 500 point and 10000 point grids and found 0.01~\% and 0.004~\% differences, respectively. Next, we investigated the construction of the finite B-spline basis set \cite{Bspline} that can accurately represent states up to $n=19$ with negligible loss of accuracy for the low-lying states. The resulting basis set consists of 150 orbitals of order 13 for each value of the relativistic angular quantum number $\kappa$ constructed in a spherical cavity of $R=600~a_0$. The quality of the basis set was verified by comparing basis set energies and dipole matrix elements with original DF values. We find 0.01~\% to 0.05~\% differences in matrix elements up to $18s-19p$, demonstrating that the basis set reproduces all of these orbitals with sufficient numerical precision. The states with $n>24$ in our basis have positive energies and provide a discrete representation of the continuum.

We use the linearized version of the coupled-cluster approach (also referred to as the all-order method), which sums infinite sets of many-body perturbation theory terms, for all  terms up to $k=40p$ in Eq.~(\ref{eq-1}). The sum over $k$ is numerically converged at $n=40$ and the estimated remainder makes a small contribution to the final uncertainty of the calculation. The details of the all-order approach are described in a review~\cite{SafJoh08}. Experimental energies from~\cite{San06,NIST_ASD} are used up to $n=19$, and theoretical energies are used for higher $n$ to keep the completeness of the basis set. The $np$ experimental energy uncertainties were listed in~\cite{San06}. The uncertainty of the $E_{18s}=33194.382(7)$~cm$^{-1}$ energy from~\cite{Sansonetti1985} was determined as the difference with the quantum defect value of 33194.389~cm$^{-1}$~\cite{SteveR}. Our theoretical removal energies are in excellent agreement with NIST values, to 0.1~cm$^{-1}$ or better for the $14p-18p$ states and 0.2~cm$^{-1}$ for the $18s$, $19p$ states serving as additional test of the theory accuracy.

Two different \textit{ab initio} all-order calculations were carried out, including only single-double (SD) excitations and additionally including the partial triple excitations (SDpT). The results for the most important transitions are listed in Table~\ref{tab1} in the SD and SDpT columns. Lowest-order DF values are listed as well for comparison. We also carried out additional calculations that included a semiempirical estimate of higher-order corrections; these results are listed as SD$_{sc}$ in Table~\ref{tab1}. SDpT \textit{ab initio} values are taken as final. The spread of the three all-order values (SD, SDpT, and SD$_{sc}$) gives the estimated uncertainty in the final value for each transition (see \cite{SafJoh08,SafSaf11} for a detailed explanation of the methodology and uncertainty evaluation). Relative uncertainty in percent is listed in the column labeled ``Unc.''. We find that the entire correlation correction, estimated as the difference of the final and DF numbers, is very small, 0.7~\% to 3~\%, further confirming the small expected uncertainties of the calculations. Relative correlation corrections are listed in the last column as percentages.
\begin{table}
\caption{Electric-dipole matrix elements (in a.u.) that give dominant contributions to the $18s$ dynamic polarizability at 1063.514~nm magic wavelength calculated in different approximations. DF values are the lowest-order Dirac Fock values. All-order single-double, scaled SD$_{sc}$, and SDpT values are listed in the corresponding columns. SDpT values are taken as final and their uncertainties are listed in column ``Unc.'' in percent. The relative correlation correction, estimated as the difference of the final and DF numbers is listed in the last column in percent.}
\label{tab1}
\begin{ruledtabular}
\begin{tabular}{lrrrrrr}
   \multicolumn{1}{c}{Transition} &
   \multicolumn{1}{c}{DF} &
  \multicolumn{1}{c}{SD} & \multicolumn{1}{c}{SD$_{sc}$} &
\multicolumn{1}{c}{SDpT} &
\multicolumn{1}{c}{Unc.} & \multicolumn{1}{c}{Corr.}\\
\hline
$18s  -  6p_{3/2}$&  0.1932&  0.1864&  0.1860&  0.1874 & 0.8~\% &   3.1~\%   \\
$18s  - 17p_{1/2}$&  178.62&  179.98&  180.66&  179.81 & 0.5~\% &   0.7~\%   \\
$18s  - 17p_{3/2}$&
257.13&  259.68&  260.62&  259.37 & 0.5~\% &   0.9~\%   \\
$18s  - 18p_{1/2}$&  222.46&  217.79&  217.63&  218.34 & 0.3~\% &   1.9~\%   \\
$18s  - 18p_{3/2}$&  310.74&  303.49&  303.13&  304.34 & 0.4~\% &   2.1~\%
 \end{tabular}
 \end{ruledtabular}
 \end{table}
 \begin{figure}[h]
\includegraphics[scale=0.46]{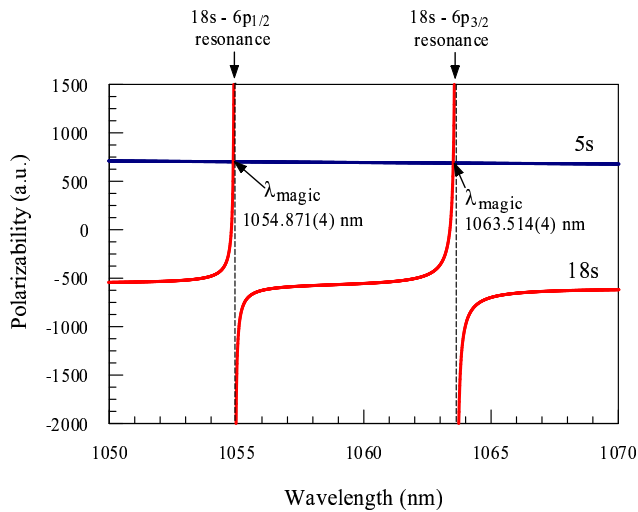}
\caption{(Color online) Magic wavelengths $\lambda$ (in nm) for the $18s$ state in rubidium, determined as crossing of the $5s$ and $18s$ dynamic polarizabilities (in a.u.).
Two magic wavelengths near the $18s - 6p_{1/2}$  and  $18s - 6p_{3/2}$ resonances are shown.}\label{fig1}
\end{figure}

\begin{table}
\caption{Contributions to the $18s$ dynamic polarizability (in a.u.) of Rb at the magic wavelength, 1063.514~nm. Reduced electric-dipole matrix elements with uncertainties listed in parenthesis (in a.u.) and the $18s-np$ energy differences (in cm$^{-1}$) are given in columns $D$ and $\Delta E$. Uncertainties in the polarizability contributions due to the uncertainties in dipole matrix elements $\delta \alpha_D$ and energies $\delta \alpha_E$ are given in the last two columns of the table, respectively, in a.u. }
\label{tab2}
\begin{ruledtabular}
\begin{tabular}{lrrrrr}
   \multicolumn{1}{c}{Contribution} &
   \multicolumn{1}{c}{$D$} &
  \multicolumn{1}{c}{$\Delta E$} & \multicolumn{1}{c}{$\alpha$} &
\multicolumn{1}{c}{$\delta \alpha_D$} &
\multicolumn{1}{c}{$\delta \alpha_E$}  \\
\hline
$(5-14)p_{1/2}$&          &              &   8.0 &  0.3  &     \\
    $15p_{1/2}$&  9.51(4) &   -223.72    &  16.8 &   0.1 &      \\
    $16p_{1/2}$& 24.70(8) &   -119.79    &  60.5 &   0.4 &      \\
    $17p_{1/2}$& 179.8(8) &   -36.84(5)  & 985.7 &  9.3  & 1.3  \\
    $18p_{1/2}$& 218.3(7) &    30.45(6)  &-1201.1& 7.9   & 2.8  \\
    $19p_{1/2}$& 25.9(1)  &     85.75    & -47.5 &  0.4  &     \\
    $20p_{1/2}$& 10.83(7) &    140.96    & -13.7 &  0.2  &     \\
    $>20p_{1/2}$&         &              & -13.8 &  6.9  &  \\ [0.5pc]
    $6p_{3/2}  $& 0.187(1)&   -9401.79(1)& 1284.4&  19.5 &   15.1   \\
$(7-13)p_{3/2} $&         &              &   17.7&    0.2&           \\
     $14p_{3/2}$&  6.72(4)&  -354.36     &  13.3 &   0.2 &           \\
     $15p_{3/2}$& 13.01(7)&  -222.19     &  31.2 &   0.3 &           \\
     $16p_{3/2}$&  34.1(1)&  -118.55     & 114.1 &  1.0  &           \\
     $17p_{3/2}$&   259(1)&   -35.84(5)  &  1995.2&  19.2&    2.8   \\
     $18p_{3/2}$&   304(1)&    31.29(6)  & -2398.0& 19.2 &   4.6    \\
     $19p_{3/2}$&  38.4(2)&    86.43     &-105.4  &1.0   &           \\
     $20p_{3/2}$&  16.4(1)&    141.74    &-31.4   &0.4   &            \\
     $21p_{3/2}$&  8.90(9)&    204.22    &-13.4   &0.3   &            \\
    $>22p_{1/2}$&         &              & -19.5 & 9.7   &            \\
Core            &         &              &  9.1  & 0.0   &            \\
Total           &         &              & 692(41)&       &
  \end{tabular}
 \end{ruledtabular}
 \end{table}

\section{Magic wavelengths}
The dynamic polarizabilities for the $18s$ and $5s$ states for wavelengths from 1050~nm to 1070~nm are shown in Fig.~\ref{fig1}. The magic wavelengths are determined as the crossing points of these two curves marked by arrows on the graph.
\begin{figure}[h]
\includegraphics[scale=0.46]{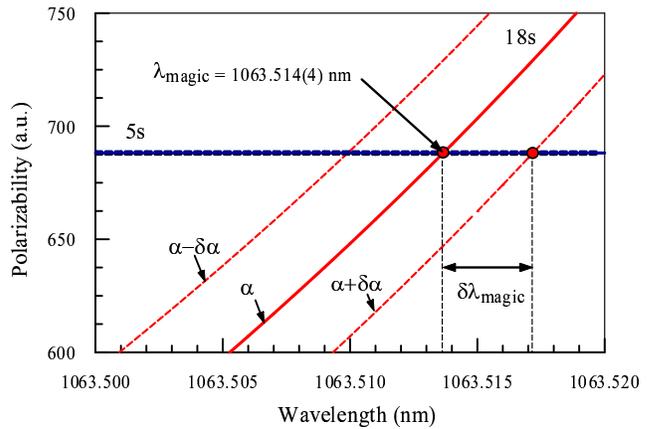}
\caption{(Color online) Determination of the uncertainty in the Rb $18s-5s$ magic wavelength.}\label{fig2}
\end{figure}
The calculation of the $5s$ polarizability has been discussed in~\cite{AroSafCla07,AroSafCla11}. The calculation of the $18s$ polarizability is illustrated in Table~\ref{tab2} where we give a breakdown of the various contribution to the $\alpha_{18s}(\omega)$ at the magic wavelength $\lambda_\textrm{magic}=1063.514$~nm.
The contributions from 15 dominant transitions are listed separately with the corresponding values of the reduced electric-dipole matrix elements and the $18s-np$ experimental energy difference in columns $D$ and $\Delta E$, respectively. The uncertainties of the dominant contributions to the polarizability arise from two sources: theoretical uncertainties in the matrix elements $D$ and uncertainties in the values of the experimental energy levels. We give these separately in columns $\delta \alpha_D$ and $\delta \alpha_E$, respectively. The latter uncertainties are only significant for 5 transitions and are omitted for all others. The uncertainties in the contributions from $np$ states with $n>20$  are (somewhat pessimistically) estimated at 50~\% based on the comparison of the lowest-order and all-order values for very highly-excited states. Adding all uncertainties in quadrature, we arrive at $\alpha_{18s}=692(41)$~a.u. The uncertainty of the $5s$ polarizability is much lower, $\alpha_{5s}=688.1(8)$~a.u., at the magic wavelength.

The uncertainties in the values of magic wavelengths are found as the maximum differences between the central values and the crossings of the $\alpha_{5s} \pm \delta \alpha_{5s}$ and $\alpha_{18s} \pm \delta \alpha_{18s}$ curves, where the $\delta \alpha$ are the uncertainties in the corresponding $5s$ and $18s$ polarizability
values. The determination of the $1063.514$~nm magic wavelength uncertainty is illustrated in Fig.~\ref{fig2}. Since the uncertainly in the value of $\alpha_{5s}$ is very small, the $\alpha_{5s} \pm \delta \alpha_{5s}$ curves blend together, and the uncertainty in the value of the magic wavelength is determined entirely by $\delta \alpha_{18s}$.

\section{Experimental procedures}
We experimentally determine the magic wavelength by measuring shifts of the $5s-18s$ two-photon transition frequency for $^{87}$Rb atoms held in a crossed-beam optical dipole trap. The trapping beams are generated by a temperature-tunable DFB laser (QD Laser QLD1061) seeding a 30~W fiber amplifier (IPG YAR-LP-SF), allowing measurements in the range from 1063~nm to 1065~nm. The wavelength is monitored with a wavemeter (WS7 High Finesse), calibrated with 40~MHz accuracy to 780~nm light stabilized to the $^{87}$Rb D2 line and verified with 1178~nm light stabilized (after frequency doubling) to the $^{23}$Na D2 line. The root-mean-square fluctuation of the measured wavelength during data taking is within the 40~MHz accuracy of the wavemeter. The dipole trap beams have a relative frequency difference of 34.8~MHz to avoid formation of a lattice. After initial cooling stages \cite{Lin2009,Brown2013}, atoms are trapped and evaporatively cooled in the crossed beams to a temperature of $\approx0.6~\mu$K, leaving a cigar-shaped cloud of $\approx6\times10^{5}$ atoms with dimensions $\approx(13\times34\times120)~\mu$m.   

Transitions to the 18s$_{1/2}$ state are driven by lasers at 795~nm (probe) and 485~nm (coupling), with an intermediate detuning between 85~MHz and 95~MHz below the $5s, F=2$ to $5p_{1/2}, F=1$ transition. Probe light is generated by a 795~nm DBR laser diode (Photodigm) narrowed to 10~kHz spectral width via polarization rotation spectroscopy in a heated vapor cell~\cite{Torii2012}. Coupling light is generated by a frequency-doubled laser at 485~nm (Toptica TA-SHG Pro) with instantaneous linewidth of less than 100~kHz, stabilized to the two-photon transition via Rydberg electromagnetically induced transparency in the same vapor cell~\cite{Abel2009}. The long-term stability of the two-photon lock is better than 0.5~MHz. The probe and coupling beams are combined on a dichroic mirror and focused along the long dimension of the atomic cloud with waists 270~$\mu$m (probe) and 170~$\mu$m (coupling). The beams have opposite circular polarizations in order to drive transitions between states of identical magnetic quantum number, reducing sensitivity to Zeeman shifts. A magnetic bias field of 0.3~mT, collinear with the excitation beams, sets the quantization axis with atoms initially in the $\left|F=2, m=-2\right\rangle$ state. 

We infer excitation to the $18s$ state by detecting 780~nm photons emitted in cascade decay through the $5p_{3/2}$ state. The light is collected by a lens relay system (NA=0.12, optimized for absorption imaging of a BEC) and focused through a 780-nm interference filter onto a 200~$\mu$m-diameter core fiber. The fiber delivers the collected light to an avalanche photodiode unit (SPCM-AQR-12) connected to a custom FPGA time-stamp card, which records arrival times of detected photons during the excitation period~\cite{Polyakov2011}. The product of geometric collection fraction, branching ratio, and detector efficiency limits the maximum probability of detection of a Rydberg excitation to less than 0.04~\%.      

For each setting of the dipole trap laser wavelength, we find the $5s-18s$ transition frequency as a function of dipole trap intensity. Following the evaporative cooling stage in the crossed dipole trap, one of the dipole beams is ramped (adiabatically to avoid heating) to a variable final intensity for the $5s-18s$ excitation. The final intensity is alternated each experimental cycle between $\approx$0.90~kW/cm$^{2}$ and $\approx$4.6~kW/cm$^{2}$. For ten wavelengths near the $6p_{3/2}-18s$ transition, the maximum dipole trap intensity is reduced to mitigate loss via the $6p_{3/2}$ state induced by the near-resonant optical trapping field. While the atoms are held in this final dipole trap, the probe frequency is scanned across two-photon resonance with a double-pass acousto-optic modulator 18 times (in alternating directions) and then the atoms are released and a new cloud loaded into the trap. We set the intensity of the probe field and the scan speed such that each scan only excites a small fraction of the atoms, but the 18 scans collectively excite nearly all of the atoms. We detect, on average, 500 photons per experimental cycle. We repeat at each dipole trap wavelength and final intensity between 5 and 10 times. 

\begin{figure}[h]
\includegraphics[width=3.5in]{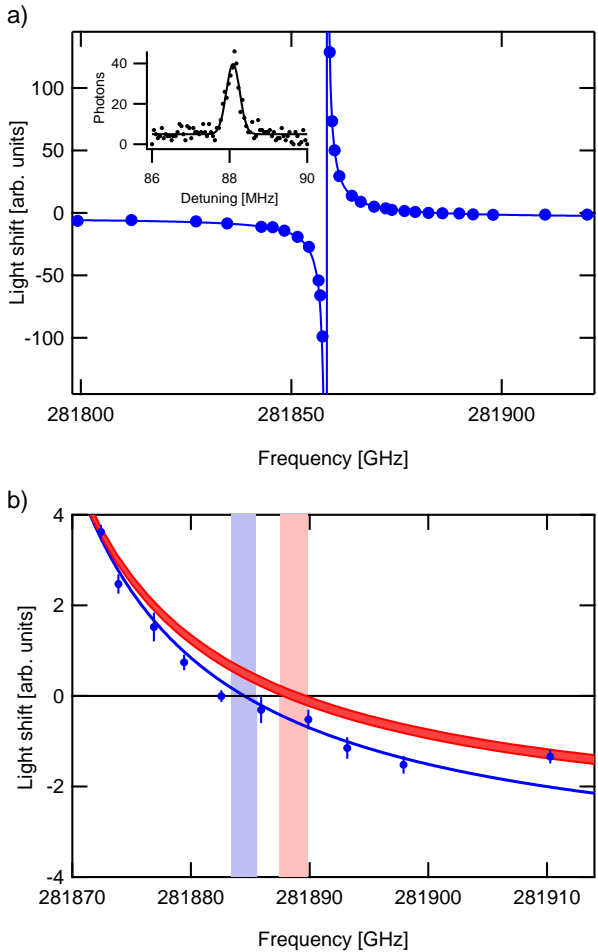}
\caption{(Color online) a) Measured $5s-18s$ light shift near the $6p_{3/2}-18s$ transition with a dispersive fit to the data. Inset is an example single-shot fluorescence spectrum with a Gaussian fit to extract the transition frequency (in terms of the probe detuning from the intermediate $5p_{1/2}$ state). b) Zoomed in on the region of the magic wavelength with a $1/x$ power law fit to the data (blue) and calculated polarizability (red) with width equal to the $18s$ polarizability uncertainty. Shaded regions are the uncertainties on the calculated (red) and experimentally extracted (blue) magic wavelengths.}\label{data}
\end{figure}

\section{Data analysis}
We correlate the arrival times of 780~nm fluorescence photons with probe detuning from the intermediate state at that time in the scan (see Fig. \ref{data}a inset for an example spectrum). We note that radiation trapping delays emission of the 780~nm fluorescence, however we measure this delay by pulsed resonant excitation of the atoms to be $\lesssim$5 $\mu$s, and set the scan speed such that the probe frequency does not change by more than one Rydberg transition linewidth in this time (scan speed $<$ (45~kHz)/(5~$\mu$s)). Following the time-to-frequency conversion, we fit a Gaussian to the emission spectrum and extract the center position of the $5s-18s$ transition (in terms of the detuning of the probe field from the intermediate $5s-5p_{1/2}$ transition). Laser frequency noise limits the width of the observed emission spectrum to $>$200~kHz and imposes a Gaussian lineshape.

For each dipole trap wavelength, we fit a line to the transition frequency as a function of dipole trap intensity. The extracted slope is linearly proportional to the differential polarizability between the ground and excited states. We plot the slope values as a function of dipole trap frequency in Fig. \ref{data}. A least-squares $1/x$ power law fit of the data on the blue side of resonance with three free parameters (zero-crossing, multiplicative constant, and additive constant) gives a zero-crossing at 281885(1)~GHz (1063.529(4)~nm), where the uncertainty is the statistical uncertainty of the fit (see Fig. \ref{data}b). The polarizability theoretically calculated above (red curve in Fig. \ref{data}b) gives a zero crossing at 1063.514(4)~nm, a difference of 4~GHz ($\approx$2.8$\sigma$). 


As an additional check of agreement with calculated level spacings, we fit the slope values around the resonance to a dispersive lineshape and extract a resonance center frequency of 281858.44(4)~GHz (see Fig. \ref{data}a). The previously measured $6p_{3/2}-18s$ transition frequency used for the all-order calculation of the magic wavelength is 281858.6(3)~GHz (see Theoretical Methods). The transition frequency extracted from our light shift measurement agrees with the theory significantly better than does the magic wavelength.

Any possible source of systematic error on the experimental measurement that could account for the discrepancy between theory and experiment requires an approximately 100~kHz shift of the $5s-18s$ transition between low and high dipole trap intensity. The van der Waals interaction between closely spaced Rydberg atoms could cause a density-dependent (and thus dipole intensity dependent) shift in the transition frequency. However, even at the highest densities and fastest excitation scan rates, we estimate a $>$2~$\mu$m average spacing between excited Rydberg atoms, which corresponds to a van der Waals interaction shift of $<$150~Hz. In addition, 100~kHz is larger than any shift we are able to measure by varying the density at a constant dipole trap intensity.

\section{Conclusion}
In summary, we have demonstrated that the relativistic all-order method is applicable to the calculation of polarizabilities of such highly-excited states as $18s$, which in addition to identifying magic wavelengths, may prove useful in precision spectroscopy measurements in the alkalis. Two convenient $5s-18s$ magic wavelengths were determined, and one experimentally verified at a wavelength accessible to commercial high-powered 1064~nm laser amplifiers.  Along with \cite{Kuzmich} and \cite{Piotrowicz2013}, our work indicates that magic trapping of Rydberg atoms is experimentally feasible over a wide range of principal quantum numbers, and should allow for increased trap lifetimes and decreased decoherence rates in a variety of Rydberg quantum information applications.

\begin{acknowledgments}
This research was performed under the sponsorship of the US Department of Commerce, National Institute of Standards and Technology and supported by the ARO's atomtronics
MURI. The identification of commercial products is for information only and does not imply recommendation or endorsement by the National Institute of Standards and Technology. E.A.G. acknowledges support from the National Research Council Research Associateship program. We thank J.~Lawall for use of his wavemeter and S.~Eckel for help with wavemeter calibration.
\end{acknowledgments}

 \bibliography{bibfile2012,exprefs}
\end{document}